\documentclass[12pt]{article}
\usepackage[utf8]{inputenc}
\usepackage{amsmath}
\usepackage[backend=biber,style=numeric,sorting=none]{biblatex}
\usepackage{graphicx}
\usepackage{caption}
\usepackage{subcaption}

\title{DEM Simulations of Spheres Flowing Through a Hopper: Validation of Beverloo Law}
\author{Leticia M. V. da Silva$^a$ \\
        \textit{Erlifas Moreira Rocha$^b$} \\
        \textit{Piter Gargarella$^b$} \\
        Pedro Augusto F. P. Moreira $^a \dagger$ \\
        $^a$\small Federal University of São Carlos, Department of Physics, \\
        \small Rodovia Washington Luís, km 235 SP-310, 13565-905 \\
        \small São Carlos, SP, Brazil\\
        $^b$\small Federal University of São Carlos, Department of Materials Engineering, \\
        \small Rodovia Washington Luís, km 235 SP-310, 13565-905 \\
        \small São Carlos, SP, Brazil\\
        $\dagger$\small \texttt{pmoreira@ufscar.br}}
\date{\today}

\begin{document}

\maketitle

\begin{abstract}
    This work presents a detailed investigation of the discharge behavior of spherical granular materials through a conical--cylindrical hopper using \emph{Discrete Element Method (DEM)} simulations. The aim is to assess the applicability limits of the empirical \emph{Beverloo law}. The system was modeled with a monodisperse particles whose mechanical properties correspond to the $Al_{95}Fe_2Cr_2Ti_1$ alloy, and interparticle contacts were described using the Hertz--Mindlin (no slip) model. The simulations systematically explored the influence of particle diameter ($d$) and bed height ($h$) on the resulting mass flow rate ($Q$).
    
    The results reveal the coexistence of transient and steady-state discharge regimes. Good agreement with the Beverloo scaling was observed for relatively small diameter ratios ($D/d = 10$) and sufficiently large bed heights, where the flow stabilizes rapidly. For larger $D/d$ ratios, the discharge rate decays exponentially, indicating a breakdown of the constant-hydrostatic-pressure assumption underlying the Beverloo model. A dimensionless criterion for the validity of the Beverloo law is proposed as $\Pi_h = h/D > 2$, or equivalently $N = h/d > 20$. The quantitative agreement between DEM simulations and experimental measurements for polydisperse particle size distributions further validates the computational model and demonstrates its predictive capability for granular discharge in confined geometries.
\end{abstract}

\section{Introduction}

Granular materials exhibit complex collective behavior that differs fundamentally from that of conventional fluids and solids \cite{nedderman1992,jaeger1996granular}. Among the most studied configurations is the discharge of particles through hoppers or silos, a process relevant to numerous industrial applications such as powder handling, metallurgy, and additive manufacturing \cite{cleary2002modelling,thornton2015granular}. Despite its apparent simplicity, granular discharge involves nonlinear interactions among particle collisions, wall friction, and stress transmission, giving rise to arching, clogging, and density fluctuations \cite{to2001,zhang2005jamming,mankoc2007flow}.

A key empirical framework to describe such flows is the Beverloo law \cite{beverloo1961}, which relates the mass flow rate $Q$ to the outlet and particle diameters through

\begin{equation}
Q = C\,\rho\,\sqrt{g}\,(D - k\,d)^{a},
\label{eq:beverloo}
\end{equation}

where $\rho$ is the particle density, $g$ the gravitational acceleration, $D$ the outlet diameter, $d$ the particle diameter, $C$ a dimensionless discharge coefficient, and $a \approx 5/2$ for three-dimensional hoppers.  The parameter $k$ is an empirical shape factor that accounts for the effective reduction of the outlet size due to the finite particle diameter, reflecting the formation of a "free-fall arch" at the orifice. This scaling assumes that the pressure near the outlet remains approximately constant, which is physically justified by the \textit{Janssen effect}. As first formulated by Janssen, the vertical stress $\sigma_z$ in a granular column saturates with depth as

\begin{equation}
\sigma_z(z) = \rho g \lambda (1 - e^{-z/\lambda}),
\end{equation}

where $\lambda = D/(4K\mu_w)$ is the characteristic screening length depending on the wall friction coefficient $\mu_w$ and the lateral-to-vertical stress ratio $K$ \cite{nedderman1992}.  
When the filling height exceeds a few times $\lambda$, the base pressure becomes nearly constant, explaining why the flow rate becomes independent of height and depends mainly on geometry and particle-scale properties \cite{aguirre2011,perge2012,mankoc2007flow,fan2023numerical}. For shallow beds or highly polydisperse systems, however, the outlet pressure varies dynamically, and the discharge rate deviates from the $(D - k d)^{5/2}$ dependence \cite{peng2021}.

Hopper geometry also influences the discharge rate: changes in cone angle, outlet edge, or wall friction can modify the effective discharge area and thus the empirical constants in Beverloo’s law \cite{cleary2002modelling,capdeville2023,wan2018influence}. Wan \textit{et al.}~\cite{wan2018influence}, for example, demonstrated that even small geometric wear at the outlet can alter the flow rate while preserving the overall Beverloo-type scaling.

The Discrete Element Method (DEM) \cite{plimpton1995fast,thornton2015granular,chen2020discrete,shire2021simulations} has become a powerful approach to examine these effects at the particle scale, resolving individual trajectories and contact forces to link microscopic parameters such as stiffness, damping, and friction to macroscopic observables like flow rate and clogging probability. Quantitative validation of DEM predictions against experiments remains essential to ensure predictive reliability, particularly for polydisperse or irregular particles \cite{fan2023numerical,capdeville2023}.

In this work, we perform large-scale DEM simulations of $Al_{95}Fe_2Cr_2Ti_1$ alloy spheres discharging through a conical--cylindrical hopper, systematically exploring the influence of particle size and size distribution on the discharge behavior. The results are analyzed in terms of transient and steady-state regimes and compared with the classical Beverloo scaling. Additionally, experimental data obtained using a gas-atomization device with controlled size distributions are used to validate the numerical predictions, providing a consistent framework for assessing the applicability limits of the Beverloo law.

\section{Computational Approach}

The simulated system consists of a three-dimensional cylindrical hopper filled with spherical particles representing granules of the $Al_{95}Fe_2Cr_2Ti_1$ alloy. The computational domain was defined as a rectangular box of dimensions $10.2 \times 10.2 \times 25.2~\mathrm{m^3}$, extending from $-5.1$ to $+5.1~\mathrm{m}$ in both $x$ and $y$ directions, and from $-10.1$ to $+15.1~\mathrm{m}$ in the vertical ($z$) direction.  

\begin{figure}
    \centering
    \includegraphics[width=0.5\linewidth]{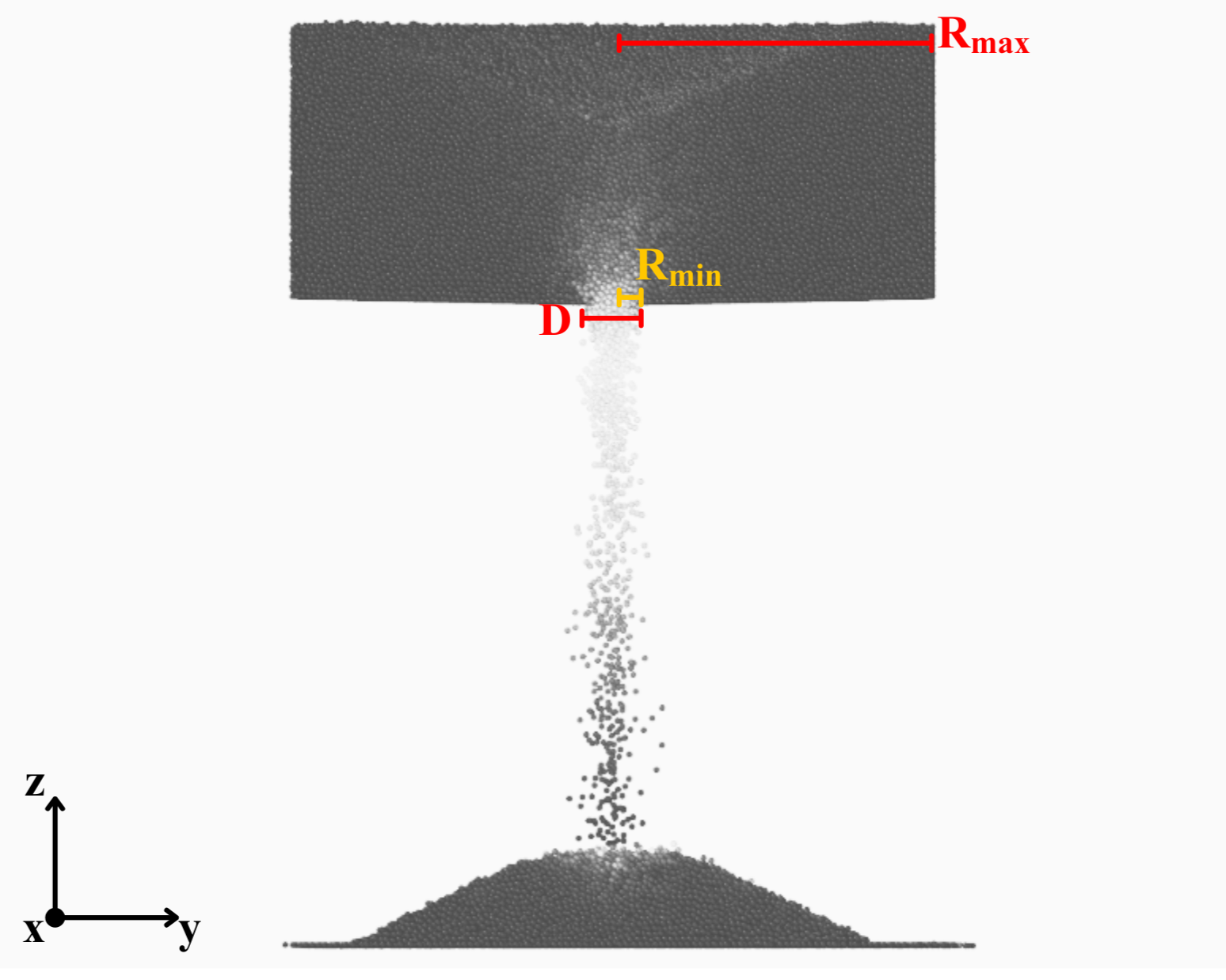}
    \caption{Representative snapshot of the granular-flow simulations through a funnel-shaped hopper. The variable $R_{max}$ denotes the maximum radius of the upper container, while $R_{min}$ corresponds to the minimum radius at the outlet. The parameter $D$ represents the diameter of the opening through which the particles exit. The coordinate axes $(x, \, y, \,z)$ indicate the orientation used in the simulations.}
    \label{fig:simulation}
\end{figure}

The hopper geometry is composed of a cylindrical upper section and a conical lower section that ends in a circular outlet. The cone connects the upper radius $R_\text{max} = 5.0~\mathrm{m}$ to the lower radius $R_\text{min} = 0.5~\mathrm{m}$, with the apex located at $z = -0.1~\mathrm{m}$. The cylindrical section extends from $z = 0$ to $z = 15~\mathrm{m}$. The outlet region, used to measure discharge rates, is a cylindrical section of height $4~\mathrm{m}$ below the cone, centered along the hopper axis. The figure \ref{fig:simulation} shows a representative snapshot of the simulations, in which the variables employed are displayed to provide a clearer understanding of the geometry used.

Seven simulations were performed, each with a different particle diameter $d$ such that the ratio between the outlet diameter and particle diameter, $D/d$, assumed the values $5$, $8$, $10$, $12.5$, $15$, $17.5$, and $20$. In all cases, the total number of particles was fixed at $N = 4\times10^5$, and the particle density was $\rho = 3000~\mathrm{kg/m^3}$. The particles were inserted randomly in the upper cylindrical region and allowed to settle under gravity before the outlet was opened, following a ``pour-and-release'' procedure. Gravitational acceleration was set to $g = 9.81~\mathrm{m/s^2}$, acting in the negative $z$-direction.

Interparticle and particle--wall contacts were modeled using the Hertz--Mindlin (no slip) model with a history-dependent tangential force, as implemented in the \textsc{LAMMPS} package \cite{plimpton1995fast}. This model accounts for normal and tangential elastic deformation, viscous damping, and static friction at the contact points.  

The normal and tangential contact stiffnesses are given by:

\begin{equation}
    k_n = \frac{4G}{3(1 - \nu)}, \qquad
    k_t = \frac{4G}{2 - \nu},
\end{equation}

where $G = E/[2(1+\nu)]$ is the shear modulus, $E$ the Young’s modulus, and $\nu$ the Poisson ratio.

For particles representing the $Al_{95}Fe_2Cr_2Ti_1$ alloy, the following material parameters were adopted:

\begin{equation}
E = 7.1\times10^7~\mathrm{Pa}, \quad \nu = 0.33, \quad \rho = 3000~\mathrm{kg/m^3}.
\end{equation}

Energy dissipation during collisions was represented by the normal and tangential damping coefficients, $\gamma_n$ and $\gamma_t$, calculated from the effective mass $m_\mathrm{eff}$ and stiffness $k_n$ as:

\begin{equation}
    \gamma_n = 10 \sqrt{\frac{4 m_\mathrm{eff} k_n}{1 + (\pi / \ln e)^2}}, \qquad 
    \gamma_t = 0.5 \, \gamma_n,
\end{equation}

where $e$ is an empirical restitution parameter. The static friction coefficient was set to $\mu = 76$, providing realistic energy dissipation and frictional effects for metallic grains. 

The same contact law was used for particle--particle and particle--wall interactions, ensuring consistent mechanical behavior across all contact interfaces. The hopper walls, bottom plate, and top cover were described through the \texttt{fix wall/gran/region} command, applying the same interaction parameters as for interparticle collisions.

Simulations were carried out using the Discrete Element Method (DEM) as implemented in \textsc{LAMMPS}. The translational and rotational equations of motion for each particle were integrated using the \texttt{nve/sphere} time integrator.  

A time step of $\Delta t = 10^{-6}~\mathrm{s}$ was adopted to ensure numerical stability. Each simulation was divided into two stages: (1) the hopper was filled during $t_\text{fill} = 10~\mathrm{s}$ with the outlet closed; (2) the bottom cap was removed, and the granular flow was allowed for $t_\text{run} = 100~\mathrm{s}$.

\section{Experimental Procedure}

The experimental validation was performed using metallic powders produced by gas atomization. A close-coupled gas atomizer, HERMIGA 75/5VI (EAC) by PSI–Phoenix Scientific Industries Ltd., was used, operating with argon as the atomization gas. The $Al_{95}Fe_2Cr_2Ti_1$ alloy was melted by induction heating at approximately 980 $^\circ$C and transferred to the atomization nozzle through a controlled-pouring mechanism. Further details about the procedure can be found in E. M. Rocha's Master thesis \cite{rocha2024tenacidade}. Upon contact with the high-pressure Argon jet (approximately 20 Bar), the molten stream disintegrated into fine droplets that rapidly solidified during free fall inside the chamber, forming nearly spherical metallic particles.

The powder was collected in two separate containers: a main vessel positioned at the base of the chamber, where the coarser fraction accumulated, and a cyclone separator connected to the exhaust line, which retained the finer fraction transported by the gas stream. The chamber operated under an inert atmosphere to prevent oxidation, and the collection efficiency between the primary vessel and the cyclone was consistent with standard
gas-atomization behavior, ensuring adequate particle recovery across all size ranges.

\begin{figure}
  \centering
  \includegraphics[width=0.75\textwidth]{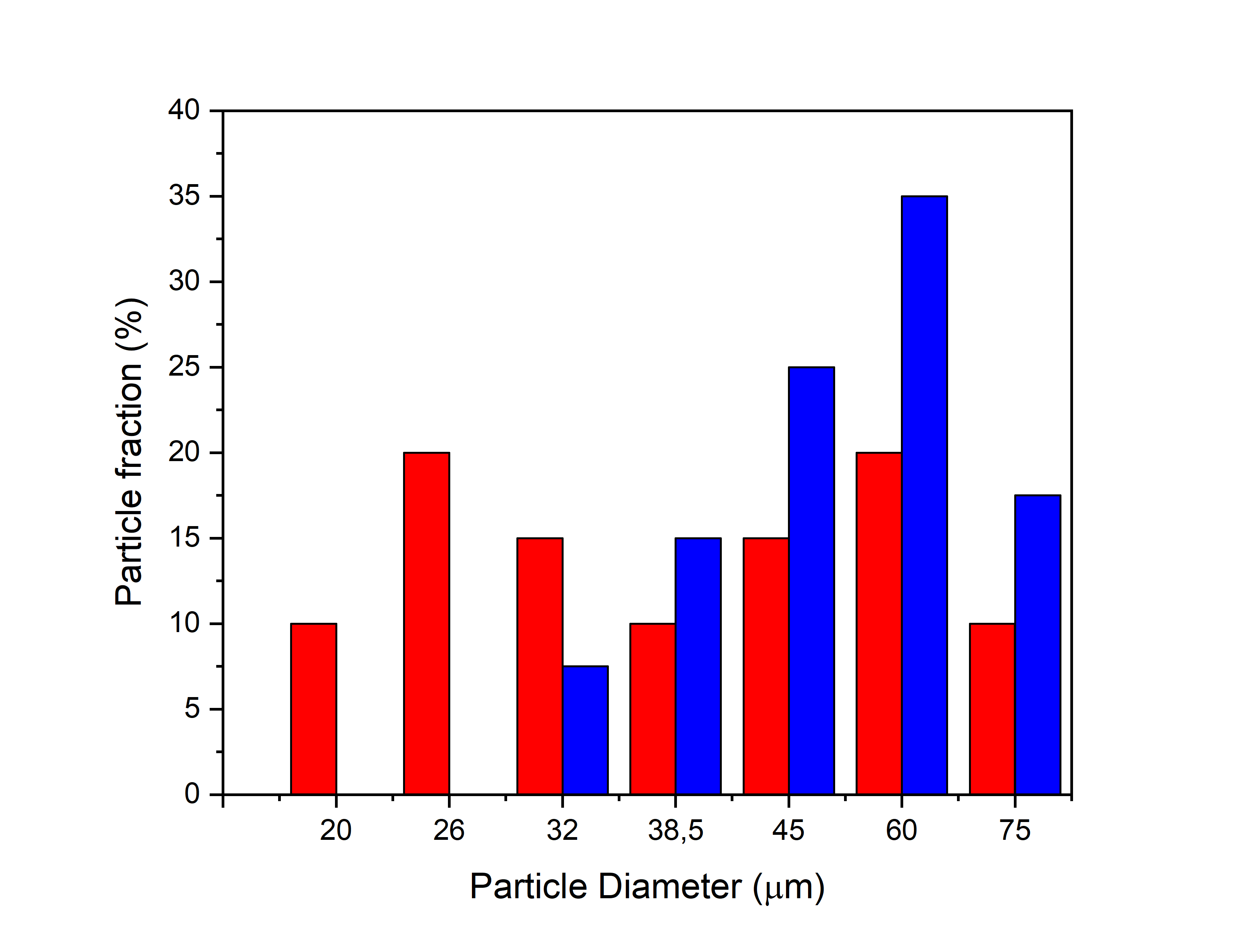}
  \caption{Particle size distributions used in the experiments:a narrow distribution (blue) and a broader one (red).}
  \label{fig:distribuicao}
\end{figure}

The gas-atomized powder was subsequently dried and sieved using a calibrated set of stainless-steel sieves with nominal openings between $20~\mu\mathrm{m}$ and $75~\mu\mathrm{m}$. The retained mass in each sieve was measured with an analytical balance, and the cumulative particle size distribution was determined from the mass fraction retained in successive sieves. Two representative distributions were selected for detailed analysis and comparison with the DEM simulations: a narrow distribution ($32$--$75~\mu\mathrm{m}$) and a broader one ($20$--$75~\mu\mathrm{m}$). The corresponding histograms of these size distributions are shown in Fig.~\ref{fig:distribuicao}.

Microscopic inspection confirmed that the particles were predominantly spherical, with smooth surfaces and no significant evidence of satellites or irregular agglomerates, supporting the assumption of spherical geometry in the numerical model. The experimentally measured size distributions were used as direct input for the DEM simulations, ensuring consistency between computational and physical conditions. The comparison between simulated and measured mass flow rates, presented in Section~\ref{results}, demonstrates the validity of the adopted modeling parameters and the applicability of Beverloo-type scaling to gas-atomized granular systems.

\section{Results and discussion}\label{results}

\begin{figure}
  \centering
  \begin{subfigure}[b]{0.45\textwidth}
    \centering
    \includegraphics[width=\textwidth]{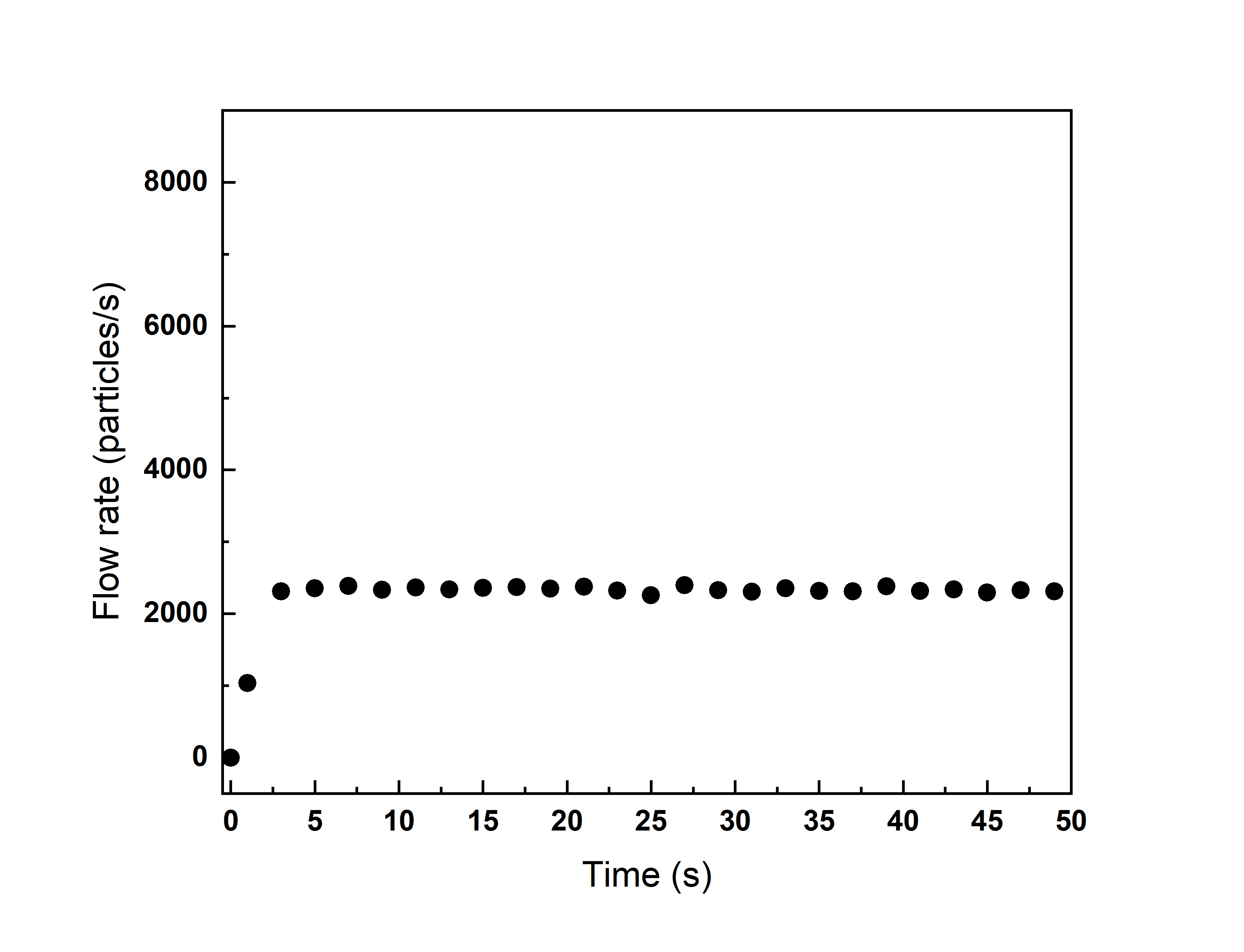}
    \caption{$D/d = 10$}
    \label{fig:1a}
  \end{subfigure}
  \hfill
  \begin{subfigure}[b]{0.45\textwidth}
    \centering
    \includegraphics[width=\textwidth]{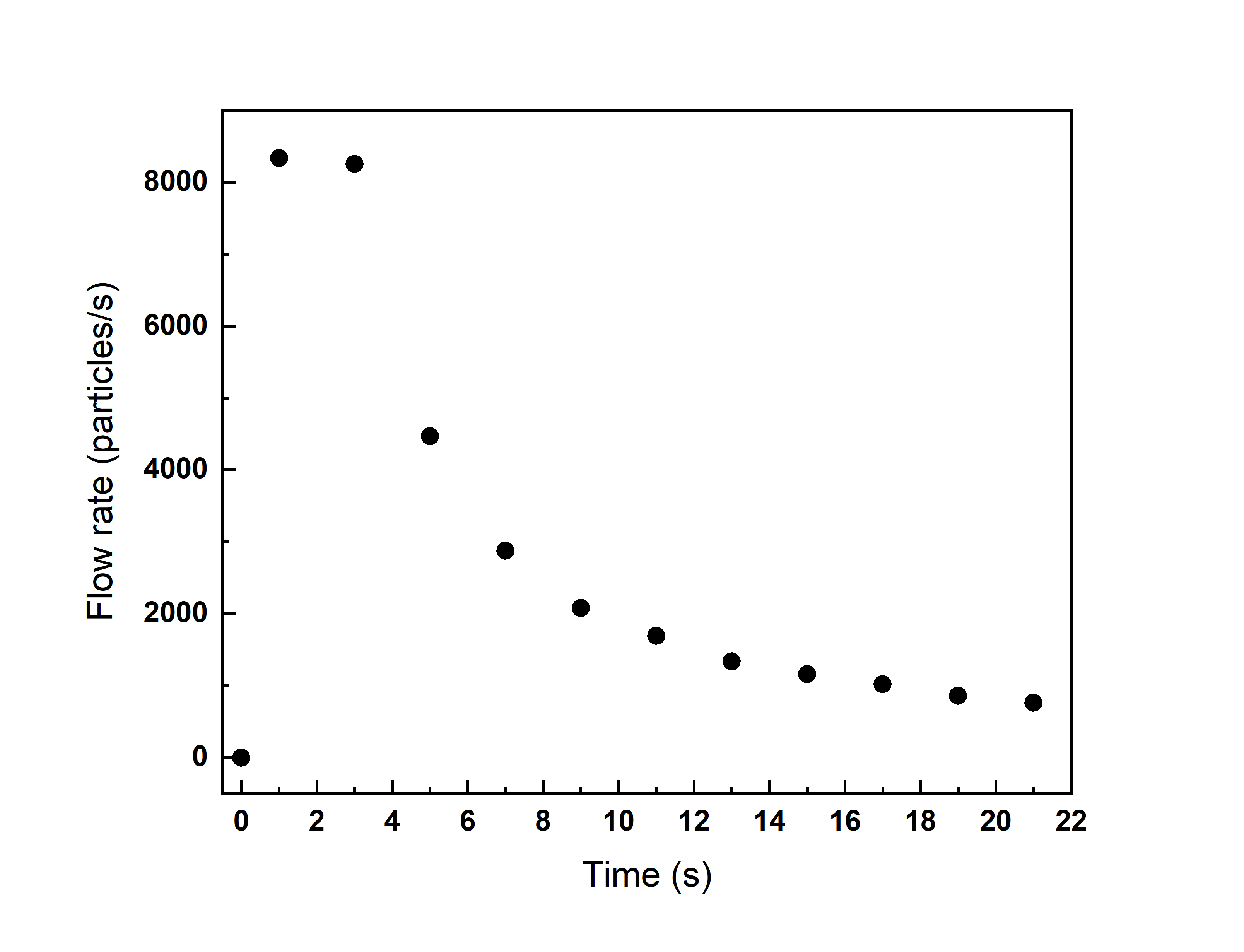}
    \caption{$D/d = 20$}
    \label{fig:1b}
  \end{subfigure}
  \caption{Flow rate as a function of time for two different ratios between the funnel diameter $D$ and particle diameter $d$.}
  \label{fig:flow_diameters}
\end{figure}

Figure~\ref{fig:flow_diameters} shows the time evolution of the particle flow rate for two hopper configurations with different diameter ratios $D/d$.

Both curves exhibit an initial transient peak corresponding to the onset of particle discharge immediately after the opening of the outlet. In the case of $D/d = 10$ [Fig.~\ref{fig:1a}], the flow rate rapidly stabilizes at a nearly constant value, indicating the establishment of a steady-state discharge regime. Conversely, for the configuration with smaller particles ($D/d = 20$) [Fig.~\ref{fig:1b}], the flow rate decreases progressively in an approximately exponential fashion, suggesting a gradual depletion of particles in the hopper. 

All simulations started with same number of particles. For the $D/d = 10$ case, the larger particle size results in a greater total mass and an initially higher bed height, which sustains a continuous pressure on the outlet and favors a steady flow. In contrast, in the $D/d = 20$ configuration the particle bed is shallower, and the lower column height $h$ leads to a reduced hydrostatic pressure, explaining the decaying discharge rate.

\begin{figure}[htbp]
  \centering
  \includegraphics[width=0.75\textwidth]{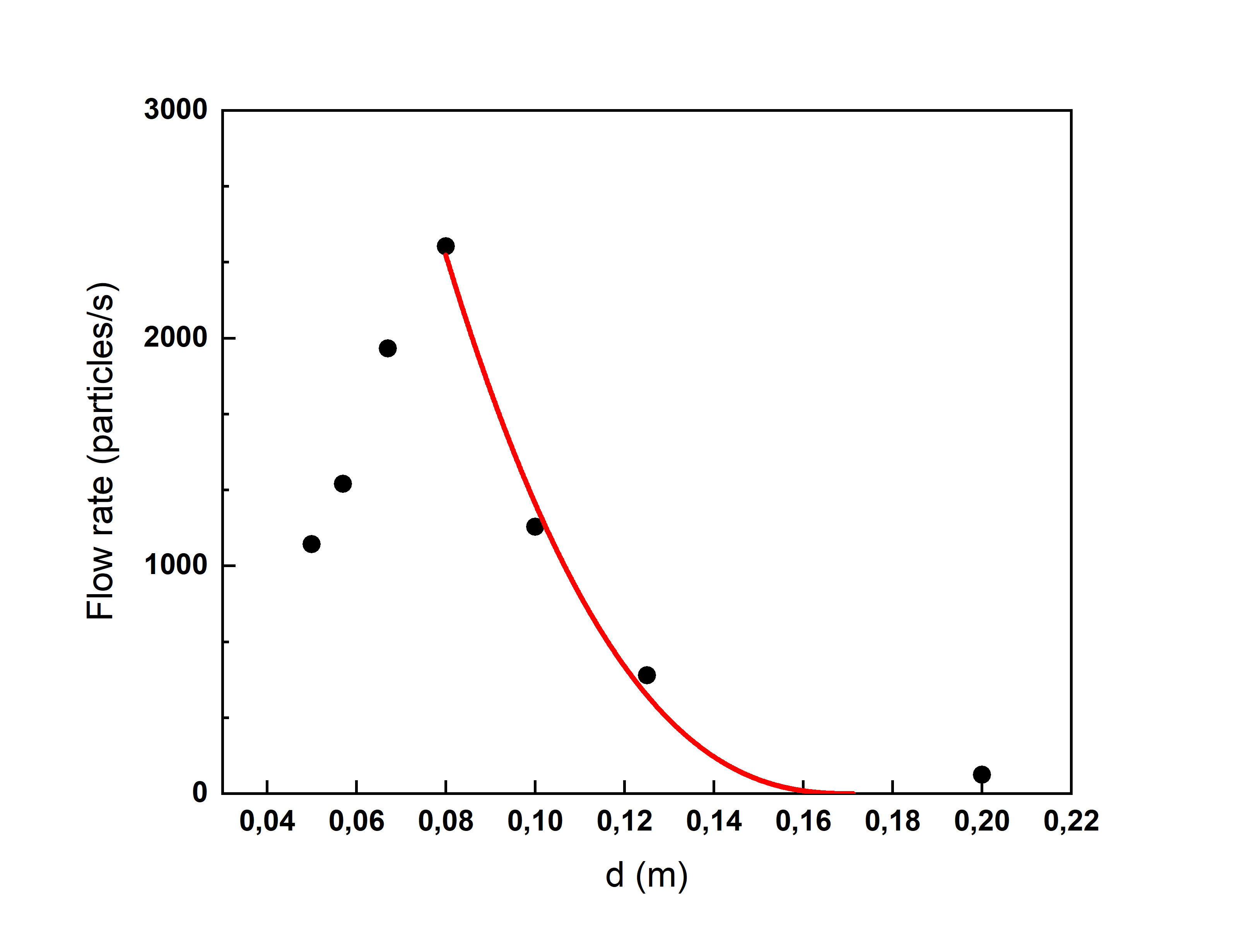}
  \caption{Flow rate as a function of particle diameter $d$ obtained from DEM simulations (black points) and compared with the Beverloo law (red line).}
  \label{fig:vazao_diametro}
\end{figure}

Figure~\ref{fig:vazao_diametro} shows the variation of the steady-state flow rate with particle diameter $d$ for a fixed funnel diameter $D$.  The black circles correspond to the DEM simulation data, and the red curve represents the fitted Beverloo-type expression, 

\begin{equation}
Q = C\,\rho\,\sqrt{g}\,(D - k\,d)^{a}.
\end{equation}

The fitting was performed by varying only the coefficient $C = 0.56$, with $a = 2.5$, $\rho = 3000~\mathrm{kg/m^3}$, and $g = 9.8~\mathrm{m/s^2}$ held fixed.  

The right-hand region of the graph, corresponding to larger particle diameters, follows closely the Beverloo law, indicating that the outlet flow is controlled by geometric constriction and that the pressure head remains approximately constant during discharge. In contrast, for smaller particles, the flow rate increases almost linearly with decreasing $d$, deviating from the $(D - k\,d)^{5/2}$ dependence predicted by the Beverloo model. This deviation occurs because the effective pressure at the outlet varies significantly as the column height diminishes, violating one of the key assumptions of the Beverloo regime.

This behavior is consistent with the time-dependent flow curves shown in Fig.~\ref{fig:flow_diameters}, where the configuration with smaller particles ($D/d = 20$) exhibited an exponential decay in flow rate,  while the case with larger particles ($D/d = 10$) quickly reached a stationary regime. Together, these results exemplify that the simulations correctly capture both the transient and steady-state regimes of granular discharge and validate the applicability limits of the Beverloo law.

\begin{figure}[htbp]
  \centering
  \includegraphics[width=0.75\textwidth]{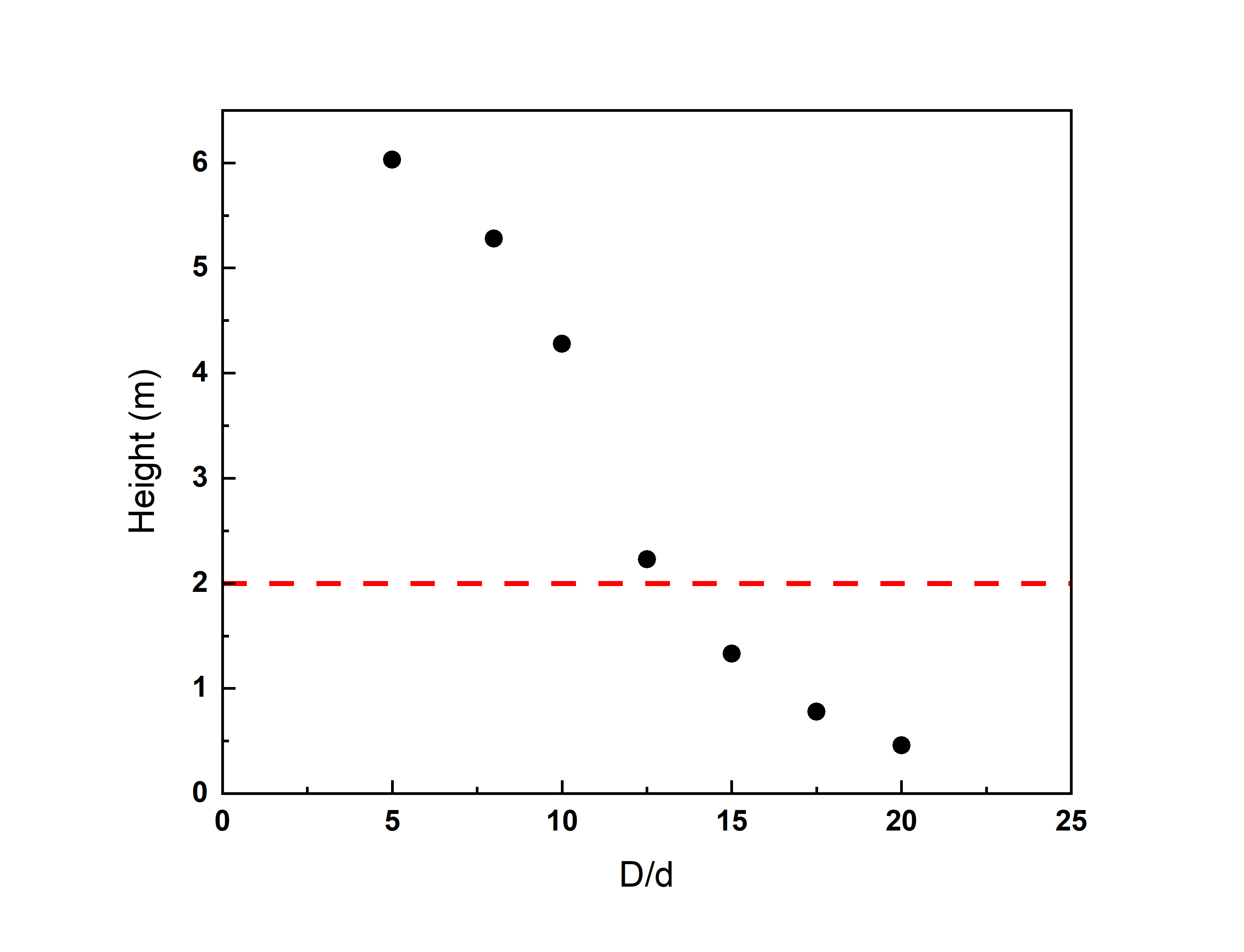}
  \caption{Initial bed height $h$ as a function of the particle diameter ratio $D/d$.  The red dashed line indicates the approximate threshold where the Beverloo law becomes valid.}
  \label{fig:altura_diametro}
\end{figure}

Figure~\ref{fig:altura_diametro} shows the initial bed height h as a function of the diameter ratio D/d, with the red dashed line indicating the approximate threshold for the Beverloo law. From these results, it is observed that the onset of Beverloo-type flow can be described by a dimensionless criterion involving the bed height, the funnel diameter, and the particle size. Two equivalent parameters can be used to express this condition: the number of particle layers $N = h/d$ and the dimensionless height $\Pi_h = h/D$. For the present system, where the transition occurs around $h = 2~\mathrm{m}$ and the typical particle diameter is $d = 0.1~\mathrm{m}$, one obtains $N_c \approx 20$ and $\Pi_{h,c} \approx 2$ as empirical thresholds. These quantities are physically related to the pressure at the outlet, which scales as $P \sim \rho_b g h$, with $\rho_b = \phi \rho$ being the bulk density of the granular packing. The Beverloo law assumes a nearly constant hydrostatic pressure during discharge, a condition that is satisfied only when the bed height is sufficiently large so that wall friction effects and stress screening do not dominate the vertical load. Considering the geometric scaling, the mass of a granular column can be approximated as $M_\text{column} \sim \rho D^2 h$, independent of the particle size $d$, reinforcing the relevance of the ratio $h/D$ as the controlling parameter. Incorporating the packing fraction $\phi$, a dimensionless pressure parameter can be defined as $\Pi_p = \phi h/D$. For $\phi \approx 0.6$ and $\Pi_{h,c} \approx 2$, this corresponds to a critical value $\Pi_{p,c} \approx 1.2$. Based on these arguments and simulation results, it is proposed that the Beverloo law is valid when $\Pi_h = h/D > 2$ or equivalently when $N = h/d > 20$, which provides a practical and general dimensionless threshold for the onset of Beverloo-type discharge in granular hopper flows.

The empirical threshold proposed here ($\Pi_h = h/D > 2$, or equivalently $N = h/d > 20$) aligns well with previous experimental and numerical observations that the mass flow rate becomes independent of the filling height once the granular column exceeds a few times the outlet diameter \cite{aguirre2011,perge2012,mankoc2007flow,fan2023numerical,peng2021, wan2018influence}. This agreement reinforces the physical relevance of the proposed dimensionless criterion as a practical indicator for the onset of Beverloo-type discharge.

\begin{figure}[htbp]
  \centering
  \includegraphics[width=0.75\textwidth]{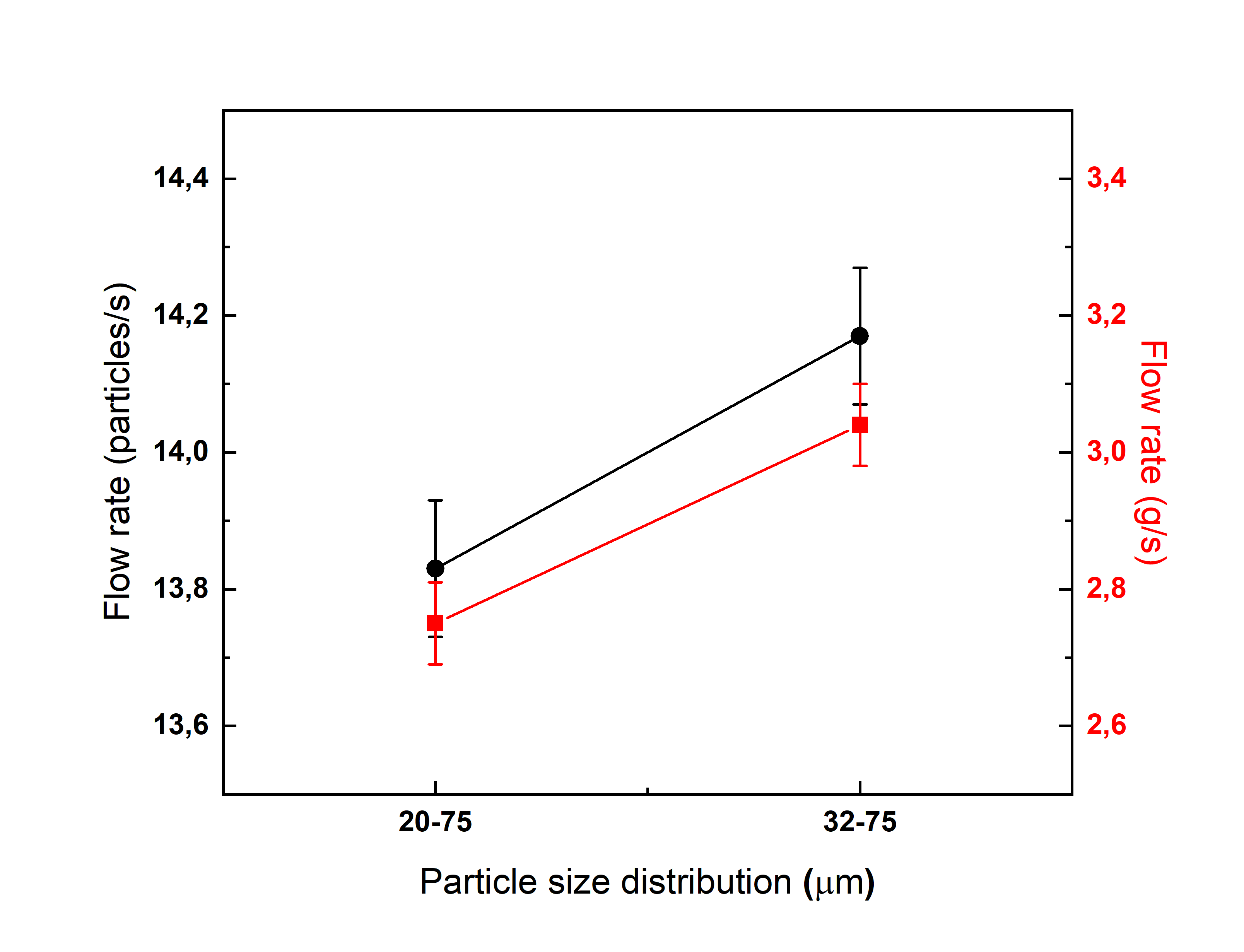}
  \caption{Comparison between experimental (red circles) and simulated (black squares) flow rates for two polydisperse particle size distributions. Error bars represent standard deviations.}
  \label{fig:vazao_distribuicao}
\end{figure}

Two polydisperse particle size distributions, similar to shown in Fig.~\ref{fig:distribuicao}, were simulated using parameters closely matching the experimental conditions described in the "Experimental Procedure" section. The geometry, particle number, and interaction parameters in the DEM simulations were chosen to replicate as accurately as possible the experimental discharge conditions, allowing a direct comparison between simulated and measured flow behaviors.

The comparison presented in Fig.~\ref{fig:vazao_distribuicao} shows that the simulated flow rates follow the same trend observed experimentally. In both cases, the broader size distribution (20–75~$\mu m$) produces a lower overall flow rate compared to the narrower one (32–75~$\mu m$), which can be attributed to enhanced packing efficiency and increased interparticle friction among smaller particles. The qualitative agreement between numerical and experimental data provides a strong validation of the computational approach used in this study. This consistency demonstrates that the discrete element model and the chosen contact parameters accurately reproduce the macroscopic discharge behavior observed experimentally, thus reinforcing the reliability of the present simulations for analyzing and predicting granular flow through hoppers with varying particle size distributions.

\section{Conclusion}

This study investigated the discharge of spherical granular materials from a conical–cylindrical hopper using Discrete Element Method (DEM) simulations, with the aim of assessing the applicability limits of the empirical Beverloo law. The simulations systematically explored the influence of particle diameter ($d$) and bed height ($h$) on the mass flow rate ($Q$).

The results showed that the validity of the Beverloo law depends strongly on the orifice-to-particle diameter ratio ($D/d$). For smaller ratios (e.g., $D/d = 10$), the flow quickly reached a steady-state regime, independent of the filling height, consistent with the assumptions of Beverloo scaling. For larger ratios (e.g., $D/d = 20$), however, the discharge rate decreased exponentially with time, indicating that the constant-pressure assumption breaks down and that the flow becomes height-dependent.

In the steady regime, the simulations confirmed that the mass flow rate follows the characteristic scaling $Q \propto (D - k d)^{5/2}$, consistent with classical empirical observations. The DEM model was also validated experimentally for polydisperse systems, successfully reproducing the observed reduction in flow rate for broader particle size distributions, attributed to higher packing efficiency and increased interparticle friction.

Overall, the results demonstrate that DEM simulations provide a robust and predictive framework for studying granular discharge. The present analysis establishes that the Beverloo law accurately describes steady, gravity-driven flows when the granular column is sufficiently tall ($\Pi_h = h/D > 2$ or $N = h/d > 20$), but fails in transient regimes where stress saturation is not achieved. These findings offer a physically grounded criterion for identifying the onset of Beverloo-type behavior and extend the understanding of flow regulation in granular hoppers.

\section*{Acknowledgments}

This research was conducted with the support of CENAPAD-SP (Centro Nacional de Processamento de Alto Desempenho em São Paulo), under the UNICAMP / FINEP – MCTI project. The authors also acknowledge the National Laboratory for Scientific Computing (LNCC/MCTI, Brazil) for providing HPC resources of the SDumont supercomputer, which contributed to the research results reported in this paper. The authors acknowledge the financial support of the Brazilian National Council for Scientific and Technological Development (CNPq) through project CNPq 444393/2024-2.

\printbibliography

\end{document}